# Stoichiometry, Spin Fluctuations, and Superconductivity in LaNiPO

T. M. McQueen[1], T. Klimczuk[2,3], A. J. Williams[1], Q. Huang[4], and R. J. Cava[1]

[1]Department of Chemistry, Princeton University Princeton NJ 08544
[2]Los Alamos National Laboratory, Los Alamos, NM 87545, USA
[3]Faculty of Applied Physics and Mathematics, Gdansk University of Technology, Narutowicza 11/12, 80-952 Gdansk, Poland
[4]NIST Center for Neutron Research, National Institute of Standards and Technology, Gaithersburg MD 20899

**Abstract**

Superconductivity in LaNiPO is disrupted by small (~5%) amounts of non-stoichiometry on the lanthanum site, even though the electronic contribution to the heat capacity increases with increasing non-stoichiometry. All samples also exhibit specific heat anomalies consistent with the presence of ferromagnetic spin fluctuations ($T_{sf} \approx 14K$). Comparison of layered nickel phosphide and nickel borocarbide superconductors reveals different structure-property correlations in the two families.

**Introduction**

Superconductors based on two dimensional $Fe_2As_2$ planes containing layers of edge-sharing Fe-As tetrahedra have received considerable attention, with $T_c$'s reaching as high as 55 $K$ [1-13]. Structurally related compounds based on nickel have also been reported[14-17], but have been the subject of fewer studies due to their lower transition temperatures and band structure calculations suggesting that the Ni variants are 'normal' electron-phonon superconductors[18, 19]. Of the known Ni-P compounds, $BaNi_2P_2$ is superconducting at 3 $K$[15], $SrNi_2P_2$ at 1.4 $K$[20], and $La_3Ni_4P_4O_2$ at 2.2 $K$[14]. LaNiPO has been reported to superconduct at 3 $K$[17] or 4.2 $K$[16], with no apparent explanation for the difference in $T_c$. Here we show that LaNiPO, containing ~4% oxygen vacancies is superconducting at 4.2 $K$, but that a La deficiency of 5% is sufficient to kill the superconductivity, despite a 25% increase in the Sommerfeld coefficient. Furthermore, we find that there is an extra contribution to the specific heat, characteristic of ferromagnetic spin fluctuations, present in all samples.

**Experimental**

Three polycrystalline samples of LaNiPO, designated "A", "B", and "C", were prepared in the same fashion. Stoichiometric quantities of dried $La_2O_3$, fresh La shavings, $Ni_5P_4$ (prepared by reaction of Ni and P by slow heating from 400 to 800 °C), and purified and dried P were pressed into pellets, placed in dry alumina crucibles, sealed in evacuated silica ampoules, and heated from 750 °C to 1050 °C over 1 hr. Each sample was then reground, repressed, placed back in an alumina crucible in an evacuated quartz tube with 2% excess P, and heated at 1200 °C overnight. After that, each sample was reground, repressed, placed in an alumina crucible in a quartz tube backfilled with 1/3 atm Ar (99.999%), and heated at 1200 °C overnight. This final step was repeated once for "A" and "B", and three times for "C". Neutron diffraction (NPD) data were collected at the NIST Center for Neutron Research on the BT-1 powder neutron diffractometer with neutrons of wavelength 1.5403 Å produced by a Cu(311) monochromator. Collimators with horizontal divergences of 15' and 20' of arc were used before and after the monochromator, and a collimator with a horizontal divergence of 7' was used after the sample. Data were collected in the $2\theta$ range of 3°–168° with a step size of 0.05°. Rietveld refinements of the structures were performed with GSAS[21] using the EXPGUI[22] interface. The neutron scattering amplitudes used in the refinements were La = 0.824, Ni = 1.030, P = 0.513 and O = 0.581 x $10^{-12}$ cm respectively. Physical property characterizations were performed on a Quantum Design Physical Properties Measurement System equipped with a $^3$He refrigerator.

**Results and Discussion**

Despite being prepared in nominally the same way, each of the three samples of LaNiPO exhibit different superconducting properties. Fig. 1 shows zero field cooled DC susceptibility data for each sample. Samples "A" and "B" both display bulk superconductivity, with transition temperatures of 4.2 $K$ and 3.2 $K$ respectively. The bulk nature of the superconductivity in samples "A" and "B" is confirmed by the presence of anomalies in the specific heat at $T_c$ (inset). In contrast, sample "C" displays only a trace of superconductivity.

Rietveld refinement of NPD data was used to understand the differences between the three samples. The structure of LaNiPO consists of $Ni_2P_2$ layers of edge-sharing Ni-P tetrahedra, separated by La-$O_2$-La units (Fig. 2(a) inset). In the unit cell, only the positions along the c axis of the lanthanum and phosphorus are not fixed by symmetry. Structural parameters were freely refined in the fits, including isotropic thermal parameters for all atoms, one positional parameter for La and P, and the occupancies of the La and O sites. The occupancies of the Ni and P sites were found not to deviate significantly from full occupancy and were therefore fixed at one for the final refinements. The refinements indicate that none of the samples are stoichiometric: the refined values in Table I are consistent with a slight (3-4%) oxygen deficiency, statistically significant at the 3.5-4σ level. While it is possible that in some circumstances reduced site occupation in structure refinements may be due to correlation of the occupancy with the thermal parameter, refinements that minimize this correlation indicate that it is real (see below). The relatively small oxygen thermal parameters, compared to phosphorus, then reflect the ionic nature of the La-$O_2$-La sandwich layer compared to the more covalent $Ni_2P_2$ layers. A typical fit is shown in Fig. 2(a), and final refinement atomic parameters and agreement values are shown in Table I. The lanthanum and phosphorus positional parameters and the oxygen site occupancies are the same for all samples within the standard deviations. However, the refined occupancy of the lanthanum site in non-superconducting sample "C" (0.951(13)) is less than that in the best superconductor, "A" (0.995(9)). This is statistically significant at the 4σ level. This slight difference in stoichiometry is not reflected in the lattice parameters due to its small amount and the competing effects of the presence of vacancies, which would tend to contract the lattice, and increased electostatics when $La^{3+}$ ions are missing, which would tend to expand the lattice. As a further confirmation that sample "C" is lanthanum deficient, and that all are oxygen deficient, a second series of refinements was performed in which occupancies were fixed at a systematic series of values near the optimal, and all other parameters were allowed to refine freely. This approach minimizes correlations between occupancy and thermal parameters. Contour plots of the fit quality factor versus La and O occupancy from such refinements in shown in Fig. 2(b) for samples "A" and "C". In both cases, the best fits are obtained with a 3-4% deficiency on the oxygen sites. While the La content of "A" is nearly stoichiometric, the best fit for "C" clearly occurs at a lower La occupancy. In addition, there is a small amount of $La_2O_3$ impurity phase (3%) in sample "C" that is not present in "A". These two observations both indicate a lower La content of the main phase in "C". Thus the structure refinements and property characterizations show that lanthanum deficiency destroys superconductivity in LaNiPO.

The normal state electronic contributions to the specific heat are also consistent with the differences in stoichiometry. The region above $T_c$ is not linear on a C/T vs. $T^2$ plot for any of the samples, even when the superconductivity has been suppressed by application of a 9 T magnetic field (Fig. 3(a) inset). Instead, there is an upturn at low temperatures, suggestive of spin fluctuations. The low temperature specific heat for a spin fluctuation system is $C = \gamma T + \beta_3 T^3 + AT^3 \ln(T/T_{sf})$[23], which includes contributions from the electrons, the lattice, and spin fluctuations respectively. Using the rules of logarithms, and dividing both sides by T, this can be rewritten in a form suitable for data fitting, $C/T = \gamma + \beta T^2 + AT^2 \ln(T)$, where $\beta = \beta_3 - A\ln(T_{sf})$. This expression fits the observed data very well (Fig. 3(a)). To fit the upturns below 1 $K$ seen for two of the high field measurements, a $B/T^3$ contribution to $C/T$, representing the high temperature part of a nuclear Schottky anomaly, was also included. A similar anomaly was observed in LaFePO[24] and LaNiAsO[25], and is ascribed to the freezing out of $^{139}$La(S=7/2) nuclear spins. The fitted values are tabulated in Table II. Under zero field, the Sommerfeld coefficient ($\gamma$) is significantly different for the three samples. The superconductor with the highest transition temperature, "A", has $\gamma = 5.1(2)\frac{mJ}{molNiK^2}$. $\gamma$ increases as $T_c$ decreases ("B", $\gamma = 5.4(2)\frac{mJ}{molNiK^2}$) and finally is largest when superconductivity disappears ("C", $\gamma = 6.5(2)\frac{mJ}{molNiK^2}$). This increase in the electronic contribution to the specific heat is consistent with greater chemical doping in "B" and "C", which apparently arises as a result of the lanthanum deficiency found in the NPD refinements (Fig. 3(b) inset). In contrast, the spin fluctuation contributions are similar for the three samples. From the definition $\beta = \beta_3 - A\ln(T_{sf})$, a plot of β versus A should be linear, with slope $-\ln(T_{sf})$ and intercept $\beta_3$, corresponding to the spin fluctuation and lattice contributions respectively. A straight line is observed for all of the fitted values of β and A (Fig. 3(b)). Furthermore, there is no substantial difference of the spin fluctuation contribution between the three samples: within error, all lie on the same line. The extracted lattice contribution is $\beta_3 = 0.278(9)\frac{mJ}{molNiK^4}$, corresponding to a Debye temperature of $\theta_D = 300K$. This Debye temperature is similar to what has been observed for LaFePO[24] and LaNiAsO[25]. The calculated spin fluctuation temperature is $T_{sf} = 14K$. This is remarkably similar to the value obtained for LaFePO ($T_{sf} = 14.3K$ [24]), and suggests that similar phenomena may be occurring in both compounds. This low value of $T_{sf}$ is consistent with the observed decrease in the Sommerfeld coefficient under an applied field (Table II, all three samples), where the field tends to suppress the mass enhancement due to the spin fluctuations[23]. Quantitative analysis of the dependence of $\gamma$ on applied field was not performed due to the large uncertainties in the values.

## Conclusion

All samples show the presence of 4-5% oxygen vacancies, but whether this non-stoichiometry is necessary for superconductivity was not addressed in the current study. It is found that 5% La vacancies are sufficient to suppress superconductivity, despite a 25% increase in the Sommerfeld coefficient. We speculate that prolonged heating in silica tubes with trace amounts of oxygen is what results in the lanthanum deficiency. The low temperature specific heat data on all three samples are consistent with the presence of spin fluctuations. Whether these spin fluctuations are related to the superconductivity cannot be determined from the present data. If these fluctuations are related to $T_c$, then some difference in $T_{sf}$ would have been expected between the superconducting ("A" and "B") and non-superconducting ("C") specimens, but no such difference was observed. However, our observation that $T_c$ is highest in the samples with the smallest $\gamma$ is unexpected in a typical BCS picture.

The empirical structure-property relationship in the $Ni_2P_2$- and $Ni_2As_2$- based materials is different than that in another structurally related family of superconductors, the nickel borocarbides, which contain isostructural $Ni_2B_2$ layers. Fig. 4 shows the observed superconducting transition temperatures of all three sets of materials as a function of the shape of the $NiX_4$ tetrahedron[14-16, 26-29]. In the $Ni_2P_2$ and $Ni_2As_2$ materials, as the across-plane X-Ni-X bond angle decreases away from 109.45° (i.e. as the tetrahedron becomes less ideal), $T_c$ increases. This is even as the nickel atoms in the square planes get further apart. This is in contrast to the nickel borocarbide superconductors[30-34], where $T_c$ goes up at the nickel-nickel distances decrease and the tetrahedra become more ideal (Fig. 4). Further study is warranted to understand this difference in relationship between structure and superconductivity in nickel-based materials. Additionally, the results in Fig. 4 suggest that stretching the $Ni_2P_2$ layers could result in a higher $T_c$ in this class of compounds.

## Acknowledgements

T. M. McQueen gratefully acknowledges support by the national science foundation graduate research fellowship program. The work at Princeton was supported by the Department of Energy, Division of Basic Energy Sciences, grant DE-FG02-98ER45706. Identification of commercial equipment in the text is not intended to imply recommendation or endorsement by the National Institute of Standards and Technology.

**Table 1**. Refined structural parameters for three samples of LaNiPO at 298 $K$ from powder neutron data. Space group *P4/nmm* (# 129). Atomic positions: **La**: 2c (1/4,1/4,z), **Ni**: 2b (3/4,1/4,1/2), **P**: 2c (1/4,1/4,z), **O**: 2a (3/4,1/4,0). Lattice parameters are in units of $Å$, and thermal parameters are in units of $10^{-2} Å^2$. All samples contain small amount of the impurity phase $LaNi_5P_3$. Sample "A" also contains <1% $La_3Ni_4P_4O_2$.

| LaNiPO | | "A" | "B" | "C" |
|---|---|---|---|---|
| | a | 4.04669(6) | 4.04768(7) | 4.04766(9) |
| | c | 8.1089(2) | 8.1104(2) | 8.1102(3) |
| **La** | z | 0.1530(2) | 0.1529(3) | 0.1529(3) |
| | $U_{iso}$ | 0.78(5) | 0.63(6) | 0.44(7) |
| | occ | 0.995(9) | 0.985(11) | 0.951(13) |
| **Ni** | $U_{iso}$ | 1.33(4) | 1.35(5) | 1.48(6) |
| **P** | z | 0.6244(4) | 0.6246(5) | 0.6247(6) |
| | $U_{iso}$ | 1.42(6) | 1.45(8) | 1.67(10) |
| **O** | $U_{iso}$ | 0.51(6) | 0.22(7) | 0.19(9) |
| | occ | 0.966(9) | 0.951(11) | 0.954(13) |
| | $\chi^2$ | 1.136 | 0.9846 | 1.134 |
| | $R_{wp}$ | 5.58% | 5.65% | 5.83% |
| | $R_p$ | 4.55% | 4.53% | 4.73% |
| | $R(F^2)$ | 8.27% | 12.49% | 15.91% |

**Table 2.** Parameters extracted from fits of the low temperature specific heat of the LaNiPO samples under applied magnetic fields. The fit is to the equation $C/T = \gamma + \beta T^2 + AT^2 \ln(T)$, where $\beta = \beta_3 - A\ln(T_{sf})$. A $B/T^3$ contribution to $C/T$, representing the high temperature part of a nuclear Schottky anomaly, was also included for the 9 T fits. The data ranges used were 0.4 to 10 $K$ for 3+ T fields, and 5.5 to 10 $K$ for fields < 3 T (to avoid the superconducting transition contribution).

|     | Field | $\gamma$ (mJ/mol*K$^2$) | $\beta$ (mJ/mol*K$^4$) | A (mJ/mol*K$^4$) | B (mJ*K/mol) |
|-----|-------|---------------|---------------|---------------|---------------|
| "A" | 0 T   | 5.1(2)        | -0.0101(15)   | 0.1140(53)    | -             |
|     | 9 T   | 4.3(1)        | 0.106(10)     | 0.0680(40)    | -             |
| "B" | 0 T   | 5.4(2)        | 0.0046(16)    | 0.1023(65)    | -             |
|     | 9 T   | 4.5(1)        | 0.1303(26)    | 0.0540(5)     | 0.05(2)       |
| "C" | 0 T   | 6.5(2)        | -0.0220(23)   | 0.1119(9)     | -             |
|     | 1 T   | 6.0(3)        | 0.0700(28)    | 0.080(11)     | -             |
|     | 3 T   | 5.96(3)       | 0.064(6)      | 0.0840(28)    | -             |
|     | 5 T   | 5.79(3)       | 0.0773(53)    | 0.0782(23)    | -             |
|     | 9 T   | 5.3(1)        | 0.0828(17)    | 0.0763(74)    | 0.091(8)      |

**Figure 1.** DC magnetization and specific heat data show that superconductivity in LaNiPO is very sensitive to small changes in composition. The best samples superconduct at 4.2 *K*. All samples contain 4-5% oxygen vacancies. Superconductivity is suppressed by a 5% lanthanum deficiency.

**Figure 2.** (a) Final NPD refinement of the sample showing the highest $T_c$, "A". (b,c) Contour plots of the refinement agreement statistic $\chi^2$ as a function of lanthanum and oxygen occupations, with the thermal parameters allowed to refine freely. The results indicate that superconducting sample "A" is stoichiometric in La, whereas sample "C", which shows little superconductivity, is La-deficient. These plots also show that both samples have a slight oxygen deficiency (3-4%). The stars show the minima in the agreement indices, indicating the best structural models, which correspond to the final refinements in Table I.

**Figure 3.** (a) The specific heat data on all three samples under various applied fields are well described by including spin fluctuations as well as electronic and lattice terms (see text). Inset shows plots of $C/T$ vs $T^2$ at 9 T. (b) A plot of $\beta$ vs A is linear, allowing extraction of both the lattice contribution and the characteristic spin fluctuation temperature (see text). All three samples show the same characteristic temperature. However, an increase in the zero field Sommerfeld coefficient correlates with both a decrease in $T_c$ and greater La deficiency (inset).

**Figure 4.** LaNiPO and the other $Ni_2P_2$-based superconductors form a single family where greater $T_c$'s correlate with less ideal (more stretched, greater Ni-Ni distance) Ni-P tetrahedra. The nickel-arsenide compounds show a similar dependence. This is in contrast to the nickel borocarbide superconductors (structurally based on $Ni_2B_2$ planes), where $T_c$ increases when the Ni-B tetrahedra become more regular (less stretched, closer Ni-Ni distance).

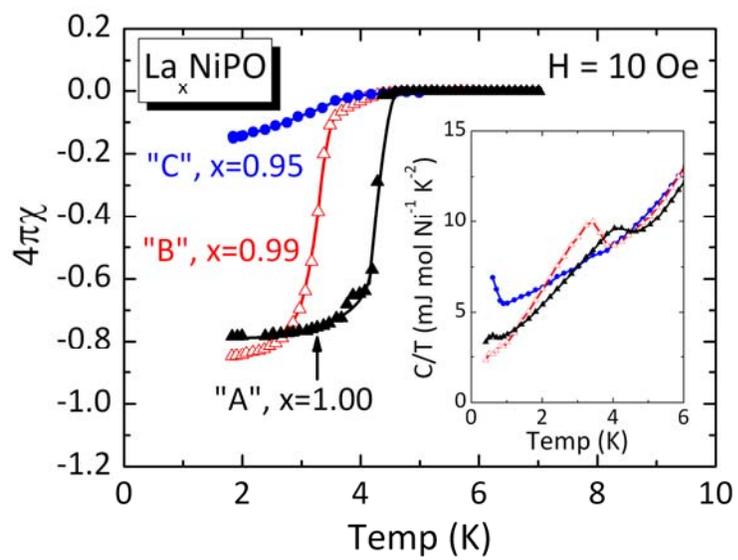

Figure 1

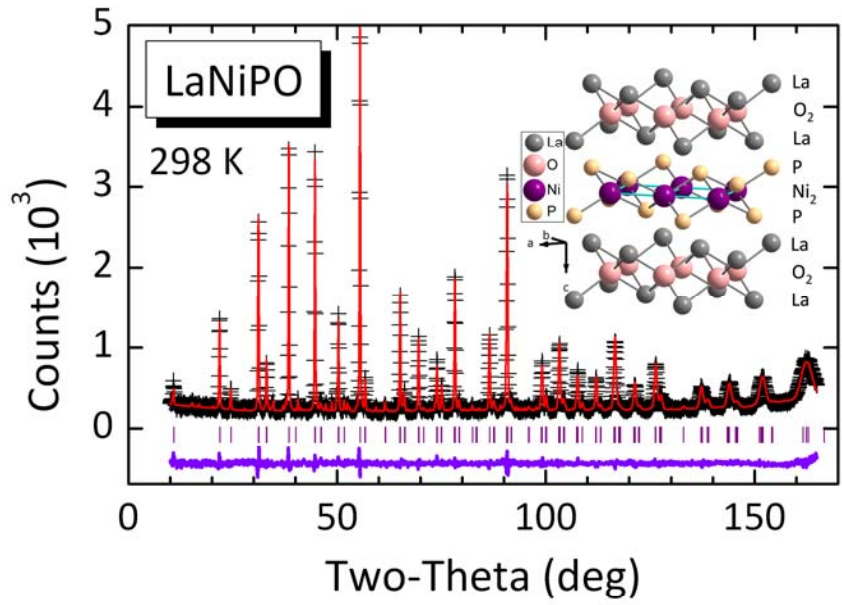
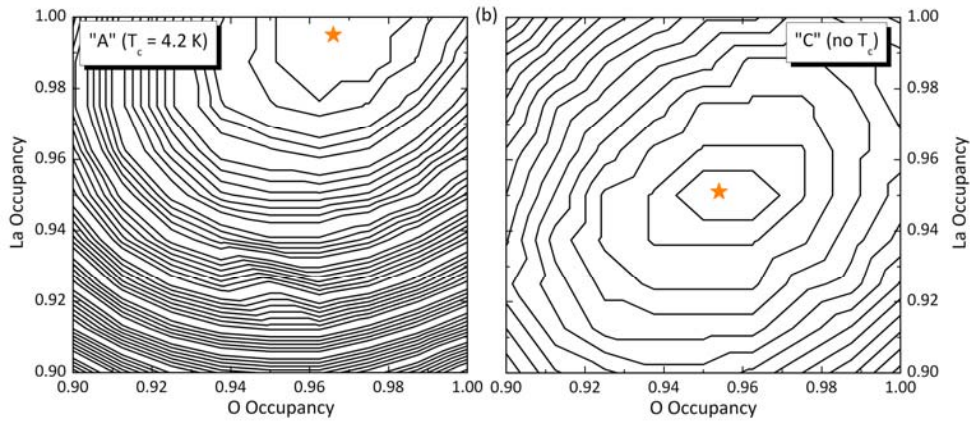

Figure 2

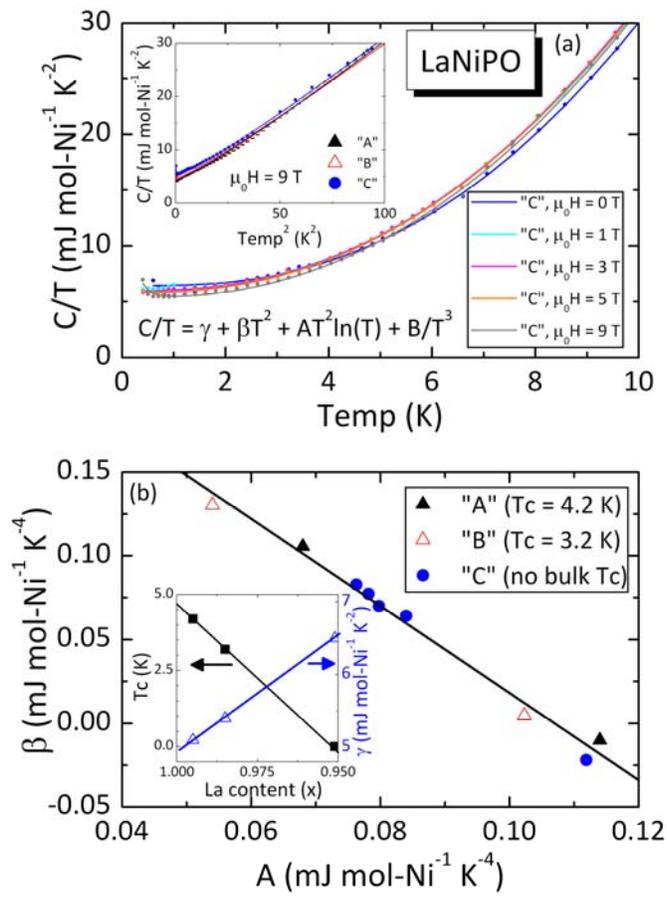

Figure 3

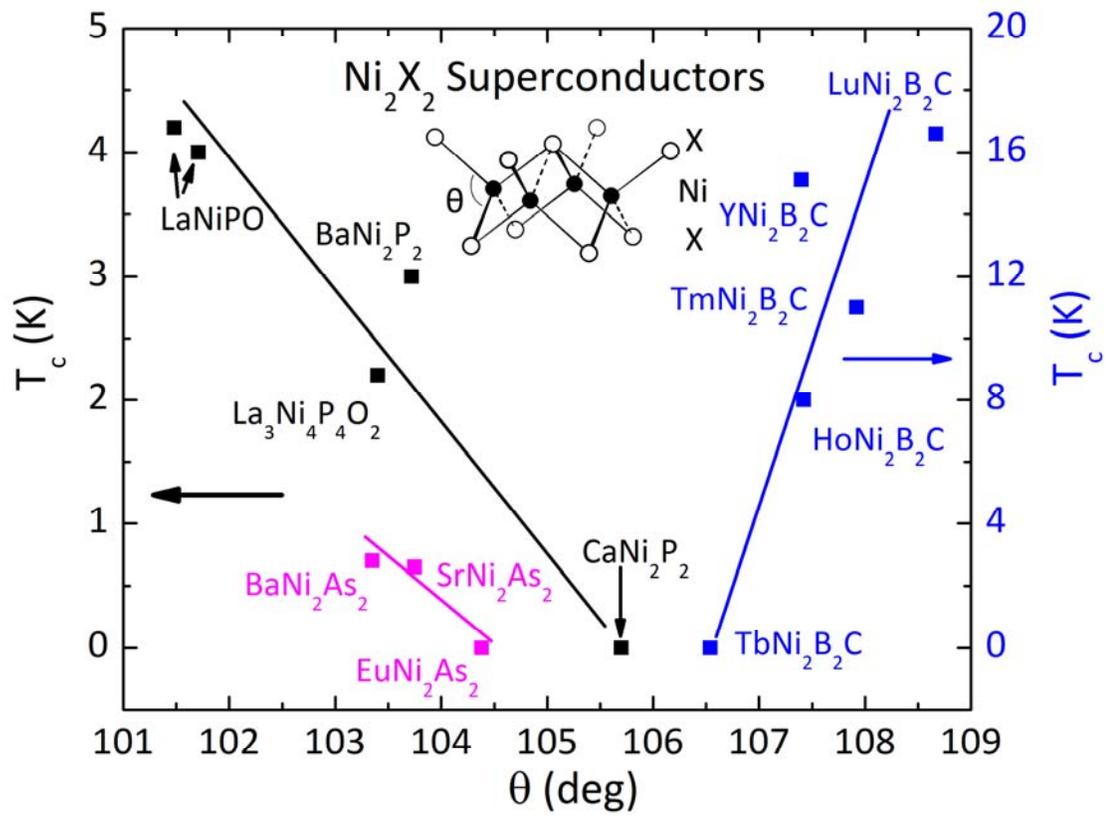

Figure 4